\documentclass[letterpaper, 10 pt, conference]{ieeeconf}  
\usepackage{mathrsfs}
\usepackage{amssymb}
\usepackage{mathtools}
\usepackage{nccmath}
\usepackage{graphicx}
\usepackage{caption}
\usepackage{pifont}
\newtheorem{lemma}{Lemma}
\newtheorem{proposition}{Proposition}
\newtheorem{definition}{Definition}

\IEEEoverridecommandlockouts                              
\title{\LARGE \bf
The Impact of Network Design Interventions on CPS Security}
\author{Pradeep Sharma Oruganti, Parinaz Naghizadeh, and Qadeer Ahmed 
\thanks{}
\thanks{Pradeep Oruganti and Qadeer Ahmed are with the Mechanical and Aerospace Engineering Department,
The Ohio State University, 
{\tt\small oruganti.6@osu.edu, ahmed.358@osu.edu}. Parinaz Naghizadeh is with the Integrated System Engineering and Electrical and Computer Engineering Departments, 
The Ohio State University,
{\tt\small naghizadeh.1@osu.edu}.}%
}

\usepackage{color}

\begin{document}

\maketitle
\thispagestyle{empty}
\pagestyle{empty}

\begin{abstract}
We study a game-theoretic model of the interactions between a Cyber-Physical System's (CPS) operator (the defender) against an attacker who launches stepping-stone attacks to reach critical assets within the CPS. We consider that, in addition to optimally allocating its security budget to protect the assets, the defender may choose to modify the CPS through network design interventions. In particular, we propose and motivate four ways in which the defender can introduce additional nodes in the CPS: these nodes may be intended as additional safeguards, be added for functional or structural redundancies, or introduce additional functionalities in the system. We analyze the security implications of each of these design interventions, and evaluate their impacts on the security of an automotive network as our case study. We motivate the choice of the attack graph for this case study and elaborate how the parameters in the resulting security game are selected using the CVSS metrics and the ISO-26262 ASIL ratings as guidance. We then use numerical experiments to verify and evaluate how our proposed network interventions may be used to guide improvements in automotive security.
\end{abstract}

\section{{Introduction}}\label{sec:intro}

Modern Cyber-Physical Systems (CPS) are increasingly targeted by attackers who use information about existing vulnerabilities in the system's components, and exploit those to launch sophisticated attacks on its safety-critical functions. A strong coupling between software and hardware components, and the (cyber-)connectivity between multiple units,  leads to opportunities for attacks that can be initiated and choreographed over multiple assets in the network, with the attackers remaining undetected for long stretches of time as they work their way to the most critical network assets~\cite{falliere2011w32,greenberg2015hackers,slay2007lessons}. As a result, CPS operators aim to optimally allocate their often limited security resources over different locations in the network to provide safety assurances. 

In this context, game-theoretic modeling and analysis can provide insights and recommendations for the CPS operators' optimal security decisions, as they can capture the contradicting goals and actions of the attackers and the defenders. There have been a number of works which have used {game theoretic} models to look at attack detection, state estimation, and optimal control of dynamic systems in presence of adversarial attacks \cite{gupta2010optimal, fawzi2014secure, pasqualetti2013attack}. There has also been significant work on security games on networks for attack detection and improving resilience \cite{pirani2021strategic, amin2013security,milani2020harnessing, nguyen2017multi,hota2018game,abdallah2020behavioral,smith2020survey, zeng2019survey}. A number of these works have used ``attack graph'' models to study attacks on inter-connected CPS. The motivation for these models is that, to successfully compromise targets internal to the network, attackers generally initiate stepping-stone attacks from external nodes, and gradually work their way to the critical assets. As such, the nodes in the attack graph are used to represent the CPS assets, while the connectivity between them shows all the components that an attacker needs to (sequentially) compromise in order to reach the CPS's most critical assets. 

In this paper, we similarly use an attack graph model to analyze how a single defender can optimally deploy its security budget to best protect a CPS against a worst-case attacker. However, contrary to these existing works, we also consider an orthogonal set of defender actions, in the form of network design interventions. In this regard, the authors have only come across \cite{milani2020harnessing} which looks at hiding or revealing edges of an attack graph to change an attacker's perception, while the original network is not modified. 

Specifically, we focus on four possible re-design actions that can be taken by a CPS operator, which result in the introduction of additional nodes in the attack graph: 
\begin{itemize}
    \item[(a)] Adding a node in series with existing nodes in the graph. Examples include adding an encryption device, or requiring stronger passwords. 
    \item[(b)] Adding a node in parallel with an existing node. Examples include adding an additional user to the CPS. 
    \item[(c)] A hybrid case of simultaneously adding a series and a parallel node to an existing node. Examples include adding an additional sensor to provide redundant information for anomaly detection. 
    \item[(d)] Adding additional input nodes. Examples include introducing an additional functionality in the system, such as adding Bluetooth connectivity to a device. 
\end{itemize} 
We will consider each of these interventions when applied to a base network. We find the equilibrium outcomes of the security game on the modified attack graph, and compare the resulting expected network losses against that of the base network to elaborate on the security implications of each design intervention. 
 
We begin by presenting the attack graph model and the security game framework in Section~\ref{sec:model}. 
Our framework is largely similar to those of \cite{hota2018game, abdallah2020behavioral}. Following this, in Section~\ref{sec:reductions}, we show that the analysis of the security game on a given attack graph $\mathcal{G}$ can be simplified by \emph{reducing} the attack graph to an equivalent attack graph $\mathcal{G}_r$. This reduction involves substituting nodes in series stretches in the graph with a single representative node, and subsequently replacing any parallel paths with common starting and ending nodes with one of the paths in that group. We show that this equivalent graph $\mathcal{G}_r$ has the same optimal investments, and the same optimal expected equilibrium loss, as the original graph $\mathcal{G}$ (Proposition~\ref{prop:net_reduction_opt}). As $\mathcal{G}_r$ can be a significantly smaller graph than $\mathcal{G}$, this simplifies the security game's analysis. 
We then outline and analyze the four network design interventions described above in Section~\ref{sec:formations}. We find that interventions (a) and (c) can lead to lower expected losses, while re-designs (b) and (d) are only justifiable if there is sufficient standalone benefits to the added nodes, as their introduction increases the expected losses of the defender at the new security game equilibrium. 

Putting together the reduction proposed in Section~\ref{sec:reductions}, with the effective re-design interventions identified in Section~\ref{sec:formations} (i.e., those that can improve the network's security), we consider the attack graph of an automotive network as our case study in Section~\ref{sec:application}. We motivate the choice of the attack graph for this case study and elaborate how the parameters in the resulting security game are selected using the CVSS metrics \cite{cvss} and the ISO-26262 ASIL ratings \cite{ISO26262} as guidance. We then use numerical experiments to verify and evaluate how our proposed network interventions can improve the vehicle's security. We discuss how the system's security is impacted by the accessibility of nodes, their role in safety-critical applications, the location in which additional safeguards are introduced in the network, and the level of information redundancy across components.  

\section{{Security Game Setup}}\label{sec:model}

This section introduces the security game framework, which includes the attack graph and the goals and capabilities of the attacker and the defender.

\subsection{The Attack Graph}
To assess the security of a given cyber-physical system, we represent its components as nodes in a directed {acyclic} graph, $\mathcal{G} = \{\mathcal{V}, \mathcal{E}\}$, where $\mathcal{V}$ represents the set of nodes and $\mathcal{E}$ represents the set of edges of the graph. A directed edge $(v_i, v_j) \in \mathcal{E}$ from node $i$ to node $j$ (i.e., $i \rightarrow j$) indicates that an attack on $j$ can be launched once $i$ is compromised. All the nodes that can be reached from a node $v \in \mathcal{V}$ (through one or more steps) are denoted as $Post(v)$, and all the nodes from which  $v$ can be reached are denoted $Pre(v)$. 
The attacks can be launched from any of the outermost \emph{entry} or \emph{source} nodes
of the graph, and are ultimately aiming to reach some \emph{target} or \emph{goal} asset. We assume that a given graph $\mathcal{G}$ can have multiple entry nodes $\mathcal{V}_s \subseteq \mathcal{V}$ from where attacks can be initiated, and a unique target node $v_g \in \mathcal{V}$. 
We use $\mathcal{P}_g$ to represent the set of paths from all sources to the target asset. Each node has an associated (financial or functional) loss $L_i \geq 0$, which is incurred if it is compromised. Further, it is assumed {that} $L_g > 0$, where $L_g$ is the loss associated with the unique target $v_g$. 

\subsection{The Security Game}

\subsubsection{Defender's actions}
To improve the security of a given system, a defender can allocate her resources to strengthen the security of its assets $\mathcal{V}$. Let $x_i\in \mathbb{R}_{\geq 0}$ denote the security investment on node $v_i$, and $\mathbf{x}=[x_1, x_2, \ldots, x_{|\mathcal{V}|}]$ denote the vector of investments on all nodes. 
We assume that given an investment $x_i$, the probability of successful attack on node $v_i$ is given by:
\begin{equation}
	p_i(x_i) = p_i^0 e^{-x_i}
\end{equation}
where $p_i^0\in [0,1]$ is the default probability that $v_i$ can be compromised in the absence of security investments. 

The defender has to choose $\mathbf{x}$ subject to a security budget $B \geq 0$, i.e., such that $\sum_{i = 1}^{{|\mathcal{V}|}} x_i \leq B$.

\subsubsection{Attacker's actions}
 
We consider a rational adversary whose action consists of selecting a path for performing stepping stone attacks. This means that the attacker selects one path in $\mathcal{P}_g$ to initiate a sequence of attacks starting from some $v \in \mathcal{V}_s$ with the goal to reach and compromise $v_g$. 

\subsubsection{The security game} The attacker chooses a path so as to maximize the expected loss inflicted on the system, i.e.,  
\begin{equation}
\mathbf{L} = \max_{P\in\mathcal{P}_g} ~ \sum_{v_i \in P}L_i  \prod_{v_j\in P\cap Pre(v_j)} p_j(x_j)~.
\label{(2)}
\end{equation}
The goal of the defender is to minimize $\mathbf{L}$ using the set of actions described above. {The attacker acts after the defender. Both players are assumed to have perfect information.} 
\section{Analyzing the Security Game through Attack Graph Reductions}\label{sec:reductions}

In this section, we present our approach for simplifying the analysis of the optimal security investment decisions for the security games of Section~\ref{sec:model}. We show that the analysis can be carried out by simplifying the attack graph through a sequence of modifications leading to a smaller attack graph with the same investment strategies and optimal costs of the starting graph. These reductions will also form the basis of our analysis of network design interventions in Section~\ref{sec:formations} and our case study in Section~\ref{sec:application}. 
Before proposing the rules for reducing a given attack graph, we formalize our notion of equivalent networks.

\begin{definition}
In the security game framework introduced in Section~\ref{sec:model}, we say two networks $\mathcal{G}_1$ and $\mathcal{G}_2$ are \emph{equivalent} if they have the same optimal investment strategies, as well as the same loss $\mathbf{L}$ under the optimal strategies. We denote this with $\mathcal{G}_1 \equiv \mathcal{G}_2$.
\end{definition}

The analysis of the optimal security investments on a general network $\mathcal{G}$ is possible by the reduction to an equivalent network $\mathcal{G}_r$, as described below. {Note that in large attack graphs where $|\mathcal{V}_r|<<|\mathcal{V}|$, this reduction can significantly reduce the number of decision variables and speed up the computation of the optimal investment profiles.}

\vspace{0.1in}

\begin{proposition}\label{prop:net_reduction_opt}
Given a network $\mathcal{G}$, construct its reduced network $\mathcal{G}_r$ by applying the following steps:
\begin{enumerate}
    \item In $\mathcal{G}$, identify stretches of series connections of the form $\{v_i \rightarrow v_{i+1} \rightarrow \ldots \rightarrow v_{i+k}\}$. Replace each with a single node $v_i'$ with loss $L_i' := \sum_{l=i}^{i+k} L_{l} \prod_{t=i}^l p^0_t$. Denote this by $\mathcal{G}'$. 
    \item In $\mathcal{G}'$, identify the set of parallel paths $\mathcal{P}_{p, i\rightarrow k}$, all originating from a common node $v_i$ and terminating in a common node $v_{k}$. Given the reduction in the previous step, these paths are of the form $\{v_i\rightarrow v_j\rightarrow v_k\}$. Replace all paths in $\mathcal{P}_{p, i\rightarrow k}$ with the path $P_j$ with $j \in \arg\max_{P\in \mathcal{P}_{p, i\rightarrow k}} p_j^0(L_j+p_k^0L_k)$. 
\end{enumerate}
Then, $\mathcal{G}\equiv \mathcal{G}_r$. 
\end{proposition}

\vspace{0.1in}

We prove this through the two Lemmas~\ref{lemma:series_loss_modification} and \ref{lemma:parallel_loss_modification}, which show that performing any of the two steps described above will result in an equivalent network. {Lemma~\ref{lemma:series_loss_modification} shows that a stretch of assets connected in series can be equivalently represented by a single asset, with an appropriately modified cost. Lemma~\ref{lemma:parallel_loss_modification} then shows how to obtain an equivalent graph by substituting a set of parallel paths with one of the paths.}

\vspace{0.1in}

\begin{lemma}\label{lemma:series_loss_modification}
Consider an attack graph $\mathcal{G}$ containing a sequence of nodes connected in series, $\{v_i \rightarrow v_{i+1} \rightarrow \ldots \rightarrow v_{i+k}\}$. Define a modified attack graph $\mathcal{G}'$, which is the same as $\mathcal{G}$, except that the path $\{v_i \rightarrow v_{i+1} \rightarrow \ldots \rightarrow v_{i+k}\}$ is replaced by a node $v_i'$ with loss $L_i'=\sum_{l=i}^{i+k} L_{l} \prod_{t=i}^l p^0_t$. Then, $\mathcal{G}\equiv \mathcal{G}'$.
\end{lemma}

\begin{lemma}\label{lemma:parallel_loss_modification}
Consider an attack graph $\mathcal{G}$ which contains a set of parallel paths $\mathcal{P}_{p, i\rightarrow k}$, all originating from a common node $v_i$ and terminating in a common node $v_{k}$. Define a modified attack graph $\mathcal{G}'$, which is the same as $\mathcal{G}$, except that all paths $\mathcal{P}_{p, i\rightarrow k}$ are replaced by the path $P_j$ with  $j \in \arg\max_{P\in \mathcal{P}_{p, i\rightarrow k}} \sum_{v_l\in P} L_{l} \prod_{v_t\in Pre(v_l)} p^0_t$. Then, $\mathcal{G}\equiv \mathcal{G}'$.
\end{lemma}

\vspace{0.1in}
It is interesting to note that these reductions resemble those used in simplifying resistor networks/circuits. Similar reductions have been observed in an alternative model for studying network security in \cite{liu2010security}. We refer interested readers to the appendix for the proofs. 
\section{Network Design Interventions}\label{sec:formations}

In the previous section, we considered attack graph reductions to find the security investment decisions of a defender protecting a \emph{given} network. Our goal in this paper is to evaluate the security implications of an orthogonal set of defender actions: interventions through network re-design. 

\begin{figure}[t]
    \centering
    \includegraphics[scale=0.2]{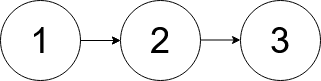}
    \caption{Base network}
    \label{base}
\vspace{0.1in}
    \captionsetup{justification=centering}
    \includegraphics[width=0.84\columnwidth]{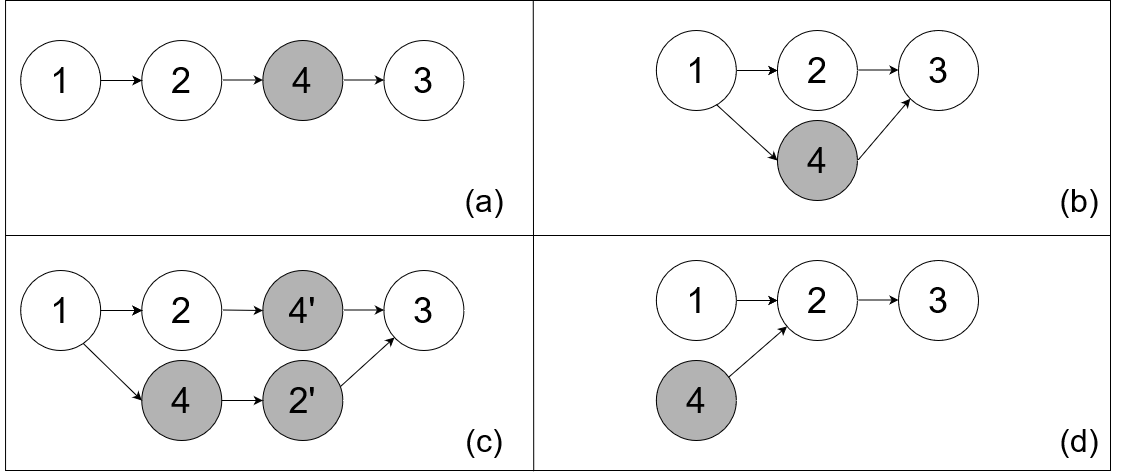}
    \caption{Network design intervention alternatives}
    \label{net_form}
\end{figure}

To this end, we begin with the base network of Figure~\ref{base}. Our base network is chosen to capture the minimal components needed for an attack graph in our context: an entry point, a target asset, and a node in between (as the target is assumed to be a key internal asset of the system which is not directly accessible from entry nodes). 
Further, by Proposition~\ref{prop:net_reduction_opt},  general attack graphs can be significantly reduced to smaller, equivalent attack graphs. 
This motivates our focus on a simple base network. 

We will motivate and assess the effects of network interventions in the four re-designed networks shown in Figure~\ref{net_form}. These networks depict possible ways of adding a new node to the base network.\footnote{Another form of design intervention may involve adding/removing links in the system. Under the current framework this would mean changing the behaviour of the system; we however assume that the goal is to re-design without removing or altering existing system functionalities. In contrast, additional nodes serve only to improve the security or introduce new functionalities in an existing system.} They are: (a) Adding a node in series; (b) Adding a node in parallel; (c) A hybrid case of networks (a) and (b); (d) Adding an additional input node. For each of these cases, we evaluate the equilibrium outcomes of the security game, and compare the resulting network losses against the base network. 

\paragraph*{A note on node's benefits vs costs} We note that each added node $v_a$ presents both a standalone loss $L_a$, as well as a potential standalone benefit $U_a$. Our analysis here focuses on the equilibrium loss $\mathbf{L}_{\text{case}}$ in each case, obtained as the solution of \eqref{(2)} (this incorporates the standalone loss $L_a$). {The difference $\mathbf{L}_{\text{case}}-\mathbf{L}_{\text{base}}$ then gives the (security) cost of performing the intervention.} If the overall loss increases by the introduction of $v_a$ (i.e., if $\mathbf{L}_{\text{case}}>\mathbf{L}_{\text{base}}$), the re-design may still be adopted as long as ${U}_a \geq \mathbf{L}_{\text{case}}-\mathbf{L}_{\text{base}}$. 

\paragraph*{The base network} The base network is shown in Figure~\ref{base}. The target node will remain the same in all formations. For ease of exposition, we let $p_i^0=p_0, \forall i,$ in this section; all results and interpretations hold for heterogeneous $p_i^0$ at the expense of more involved notation. 

By Lemma~\ref{lemma:series_loss_modification}, given a security budget $B$, the optimal security investment strategy in this case is $\mathbf{x}^* = [B, 0, 0]$, i.e., all of the investment goes on the initial node so as to achieve the maximum security benefit throughout the attack path. The loss under the optimal investment on this network is therefore $\mathbf{L}_{\text{base}} = (L_1p_0+L_2p_0^2+L_3p_0^3)e^{-B}$. 

\subsection{Series connection: increased endurance}
We begin with the case of an additional node, Node-4, added in series with existing nodes after the entry node. 
This represents a case where the defender has introduced additional safeguards in the network, such as an additional encryption device, or adding a stronger password requirement. As such, the attacker now has to compromise an extra node to get to $v_g$ along a given attack path. 

Similar to the base network, this formation consists of only one attack path, with all nodes in series. 
Therefore, by Lemma~\ref{lemma:series_loss_modification}, given budget $B$, the optimal investment strategy for this network is $\mathbf{x}=[B, 0, 0, 0]$, resulting in a loss $\mathbf{L}_{\text{series}}= (L_1p_0+L_2p_0^2+L_4p_0^3+L_3p_0^4)e^{-B}$. 

We conclude that while a node added in series over a given attack path does not change the optimal defense strategy, it does affect the cost incurred from a successful attack on the target. In particular, nodes added in series can be interpreted as methods to increase the \emph{endurance} of the system: denoting the additional node by $v_a$, their introduction reduces the probability of successful attack on all nodes in $Post(v_a)$ by a factor $p_a(x_a)$. When the standalone loss $L_a$ of $v_a$ is sufficiently small, this means that $\mathbf{L}_{\text{series}}\leq \mathbf{L}_{\text{base}}$.

\subsection{Parallel connection: structural redundancy}
We next introduce the additional Node-4 in parallel with Node-2. This may be viewed as a network in which there is \emph{structural redundancy}. A structurally redundant path captures situations where completely independent assets perform the same function. Although a security compromise of one of the parallel nodes does not lead to complete system failure, the attacker still has the opportunity to launch an attack on the subsequent nodes. As an example, nodes 2 and 4 can be interpreted as different users of the CPS. 

Here, by Lemma~\ref{lemma:parallel_loss_modification}, given the security budget $B$, the optimal investment strategy is $\mathbf{x}^*=[B, 0, 0, 0]$, indicating that the full budget should be deployed on the entry node. The loss under this profile is therefore $\mathbf{L}_{\text{parallel}}=\max \{(L_1p_0+L_2p_0^2+L_3p_0^3), (L_1p_0+L_4p_0^2+L_3p_0^3)\}e^{-B}$. 

This means that, despite not affecting the optimal security investment strategy, the parallel node could impact the defender's incurred loss. In particular, the equilibrium loss may either increase or remain the same, depending on if the additional node has a higher or lower standalone loss compared to the node it is parallel with.   

\subsection{Hybrid connection: functional redundancy}
Next, we consider a hybrid of the series and parallel cases. This network re-design is introduced so as to assess the impact of introducing \emph{functional redundancy} in a system. A functionally redundant component can be used to carry out the same tasks as an existing node; for instance an additional sensor can be added to attain signals for health monitoring, or anomaly detection and isolation. While functioning independently, information from such components is generally used in unison for decision making. 

Typically, redundancy would be represented as a parallel path similar to case (b). However, given that input from both components are ultimately used in unison for decision making by subsequent components, a successful attacker would need to compromise \emph{both} nodes (as least to some extent) to proceed in the stepping stone attack towards $v_g$. This consideration is incorporated in this network formation as the additional nodes, Node-2' and Node-4', following the parallel nodes. We assume these nodes do not have any standalone loss associated with them, and that the probability of successful attack on each node is $p_0$. 

Following the reduction in Proposition~\ref{prop:net_reduction_opt}, the optimal investment strategy for this network is $\mathbf{x}^*= [B, 0, 0, 0]$, leading to defender's loss  $\mathbf{L}_{\text{hybrid}}=\max\{(L_1p_0 + L_2p_0^2 + L_3p_0^4), (L_1p_0 + L_4p_0^2 + L_3p_0^4)\}e^{-B}$. 

Like the parallel connection, the optimal investment strategy remains the same while the equilibrium cost depends on the losses over the newly introduced attack path. Additionally, similar to the series connection, the probability of successful attack on all nodes in $Post(v_a)$ is reduced by a factor $p_a(x_a)$. Compared to the base network, the optimal loss depends on the security characteristics of $v_a$ and/or $v_g$: if the added $v_a$ has at most the same loss as its parallel counterpart, or if $L_g$ sufficiently large, then $\mathbf{L}_{\text{hybrid}} < \mathbf{L}_{\text{base}}$. 

\subsection{Additional input: new functionalities}
Lastly, we consider the additional node, Node-4, appearing as an added input node. This case captures network modifications that introduce \emph{additional functionalities} which can lead to an increased attack surface. For example, adding Wi-Fi access in addition to Bluetooth may increase the functionality of a component, but adds new vulnerabilities. 

Compared to the base network, the set of input nodes is now $\mathcal{V}_s = \{1, 4\}$, creating an additional attack path that leads to $v_3$. 
The defender's objective is given by
\begin{equation*}
\begin{split}
    \min_\mathbf{x} \max \{ (L_1p_0e^{-x_1} + L_2p_0^2e^{-x_1-x_2} + L_3p_0^3e^{-x_1-x_2-x_3}), \\
    (L_4p_0e^{-x_4} + L_2p_0^2e^{-x_4-x_2} + L_3p_0^3e^{-x_4-x_2-x_3})\}
\end{split}
\end{equation*}
Assuming that the initial nodes have the same standalone loss $L_1=L_4=L$, the optimal investment strategy subject to budget $B$ is $\mathbf{x}^* = \{0.5(B+ \log(\frac{L}{p_0L_2+p_0^2L_3})),- \log(\frac{L}{p_0L_2+p_0^2L_3}), 0, 0.5(B+ \log(\frac{L}{p_0L_2+p_0^2L_3}))\}$ when $L\leq p_0L_2+p_0^2L_3$, and is $\mathbf{x}^* = \{0.5B, 0, 0.5B,0\}$ otherwise. Consider the former case; this is when the standalone loss from the entry nodes is not higher than that of subsequent nodes combined. The loss under this investment is $\mathbf{L}_{\text{add-input}} = 2p_0e^{-B/2}\sqrt{p_0LL_2+p_0^2LL_3}$. 

For this case, we observe that having additional inputs can significantly impact the optimal investment strategy when compared to the previous networks. Specifically, in addition to investments on the two input nodes (as done in prior cases, too), a significant investment can also be placed on Node-2. In addition, the defender's equilibrium loss is higher in this case, i.e. $\mathbf{L}_{\text{add-input}}>\mathbf{L}_{\text{base}}$; this is consistent with the expectation that adding components that increase that attack surface make the system harder to secure. 
\section{Applications in Automotive Security}\label{sec:application} 

In this section, our proposed methodology is applied onto the analysis of interventions in the network of an automotive system. We begin by describing our proposed attack graph for this case study, illustrated in Figure~\ref{fig: basic_atk_graph}, and elaborate how the parameters in the resulting security game are selected using the CVSS metrics \cite{cvss} and the ISO-26262 ASIL ratings \cite{ISO26262} as guidance. We then use numerical experiments to verify and evaluate how our proposed network interventions may improve the vehicle's security.

\subsection{The Attack Graph: Accessing a Vehicle's CAN through the IVI module} 
The attack surface for a Connected and Autonomous Vehicle (CAV) is quite large. A typical vehicle consists of close to 100 Electronic Control Units (ECUs), connected via multiple different protocols with varying degrees of security. Additionally, many of the internal components are designed by multiple suppliers and vendors.  

Over the past decade, multiple studies have successfully demonstrated attacks targeting automotive systems' safety-critical functions \cite{checkoway2011comprehensive,koscher2010experimental}. These attacks typically involve the attackers reaching the vehicle's internal Controller Area Network (CAN).  
Moreover, the increasing implementations of connectivity features on the vehicles increase the number of entry points for attackers, who can remotely gain access and manipulate messages on the CAN bus; one such attack was famously demonstrated in \cite{greenberg2015hackers}. 
We therefore consider the CAN as the target node, or crown-jewel, of the attack graph. 

\begin{figure}[t]
    \centering
    \captionsetup{justification=centering}
    \includegraphics[width=0.54\columnwidth]{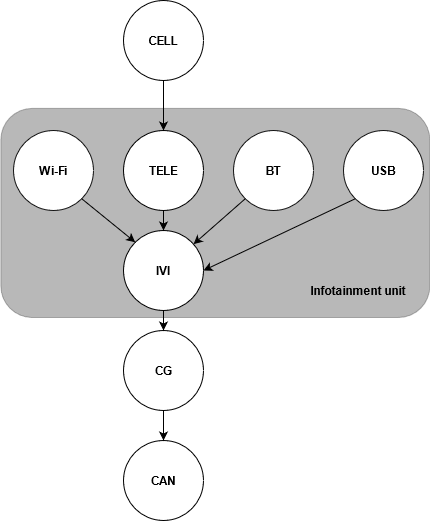}
    \caption{Attack graph for CAN access via. the In-Vehicle-Infotainment (IVI) module}
    \label{fig: basic_atk_graph}
    \vspace{-0.12in}
\end{figure}

We consider remote attacks carried out through the In-Vehicle Infotainment (IVI) system. A typical IVI module provides local connectivity options \textit{viz.} Wi-Fi, bluetooth (BT), and USB. Additionally if present, the telematics component (TELE) is connected to an external cellular communications (CELL) and GPS for general connectivity, location, and remote update services. Further, via the gateway unit (CG), the IVI unit is connected to the internal CAN lines connected to ECUs which may provide Advanced Driver Assistance (ADAS) or other chassis ECUs. An attacker can thus trigger a safety hazard by remotely gaining access to this network by compromising the IVI module \cite{moller2019guide, kawanishi2019comparative}. A basic attack graph $\mathcal{G}$ for such an attack is shown in Figure~\ref{fig: basic_atk_graph}. 

\subsection{Setting the Security Game Parameters}
In this paper, we use the CVSS metrics \cite{cvss} and the ISO-26262 ASIL ratings \cite{ISO26262} as guidance to assign values to $p^0_i$ and $L_i$, $i \in \mathcal{V}$. To quantify $p^0_i$, we look at a node's Exploitability (EX) which is obtained from its Attack Vector (AV), Attack Complexity (AC), Privileges Required (PR), and User Interaction (UI) metrics. The Automotive Safety \& Integrity Levels (ASIL) are used to gauge the relative criticality of the components in-view of safety with a higher rated ($\text{ASIL-A}< \text{ASIL-B}< \text{ASIL-C}< \text{ASIL-D}$) node receiving a higher loss. The derived $p_0$ and $L$ values associated with each node are shown in Tables \ref{table: p0_nodes}. Additional details are given in the appendix. 

\begin{table}
\caption{Probability of attack $p_0$ (translated from CVSSv3 exploitability (EX) metric) and Loss $L$ associated with each node (based on safety-criticality)}
\label{table: p0_nodes}
\begin{center}
\begin{tabular}{|c|@{}p{4pt}@{}|c||c||c||c||c||c|@{}p{4pt}@{}|c|}
\cline{1-1} \cline{3-8} \cline{10-10}
Node && AV & AC & PR & UI & EX & $p^0$ && $L$\\
\cline{1-1} \cline{3-8}\cline{10-10}
CELL && N & H & L & N & 1.62 & 0.162 && 1\\
\cline{1-1} \cline{3-8}\cline{10-10}
TELE && A & H & L & N & 1.19 & 0.119 && 10\\
\cline{1-1} \cline{3-8}\cline{10-10}
WiFi && A & L & L & N & 2.07 & 0.207 && 1\\
\cline{1-1} \cline{3-8}\cline{10-10}
BT && A & L & L & N & 2.07 & 0.207 && 1\\
\cline{1-1} \cline{3-8}\cline{10-10}
USB && P & L & L & N & 0.67 & 0.067 && 1\\
\cline{1-1} \cline{3-8}\cline{10-10}
IVI && A & H & L & N & 1.19 & 0.119 && 10\\
\cline{1-1} \cline{3-8}\cline{10-10}
CG && A & H & L & N & 1.19 & 0.119 && 50 \\
\cline{1-1} \cline{3-8}\cline{10-10}
CAN && A & L & L & N & 2.07 & 0.207 && 100\\
\cline{1-1} \cline{3-8}\cline{10-10}
\end{tabular}
\end{center}
\end{table}

\subsection{Effects of Design Interventions on the Vehicle's Security}
\begin{figure}[t]
    \centering
    \captionsetup{justification=centering}
    \includegraphics[width=0.54\columnwidth]{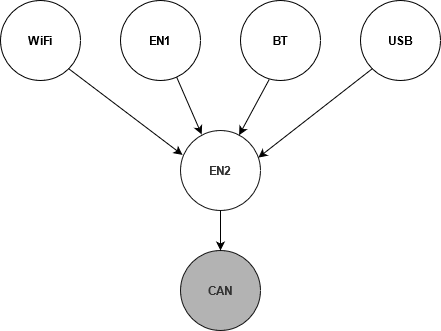}
    \caption{Reduced attack graph of attack shown in Fig. \ref{fig: basic_atk_graph}.}
    \vspace{-0.1in}
    \label{reduced_attack_graph}
\end{figure}

We first apply \eqref{(2)} directly to obtain the expected loss for the system.  It is assumed that the designer has a budget $B$ of 10 units. The optimal investment profile  is $\mathbf{x^*} = [2.53,0,2.36, 2.36,1.23,1.52,0,0]$, leading to an optimal loss, $\mathbf{L_{\text{base}}}$ of 0.0289 units. Next, applying Proposition~\ref{prop:net_reduction_opt}, we get the reduced graph, $\mathcal{G}_r$ for this system, shown in Figure~\ref{reduced_attack_graph}. {The replaced equivalent nodes EN1 and EN2 represent the CELL and IVI nodes respectively with the modified loss terms following Lemma~\ref{lemma:series_loss_modification}.}
The target node $v_g$ is grayed out as there is no investment on this node. 
The expected loss for $\mathcal{G}_r$, calculated from \eqref{(2)}, and the corresponding optimal investments, comes out to be the same as $\mathcal{G}$, concluding that $\mathcal{G}_r \equiv \mathcal{G}$. 
The number of decision variables is reduced from 8 in $\mathcal{G}$ to 5 in $\mathcal{G}_r$. 

\subsubsection{Impacts of the entry nodes (functionalities)} 
\begin{figure}[t]
    \centering
    \captionsetup{justification=centering}
    \includegraphics[width=0.7\columnwidth]{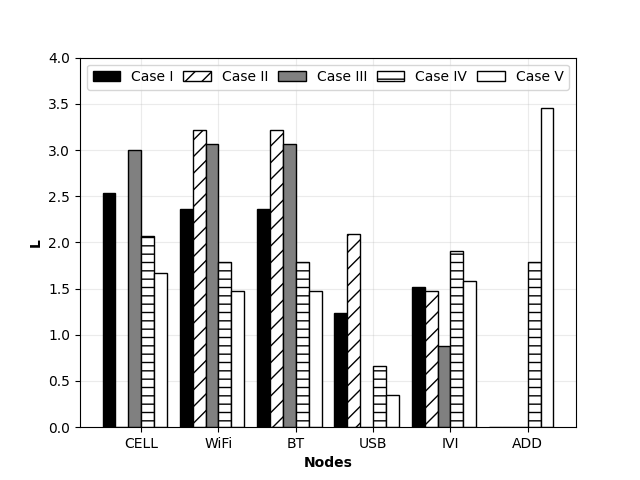}
    \caption{Security investment on each asset when input nodes change, for different configurations of entry nodes.}
    \label{fig: bar_inp_nodes}
\end{figure}

\begin{figure}[t]
    \centering
    \captionsetup{justification=centering}
    \includegraphics[width=0.6\columnwidth]{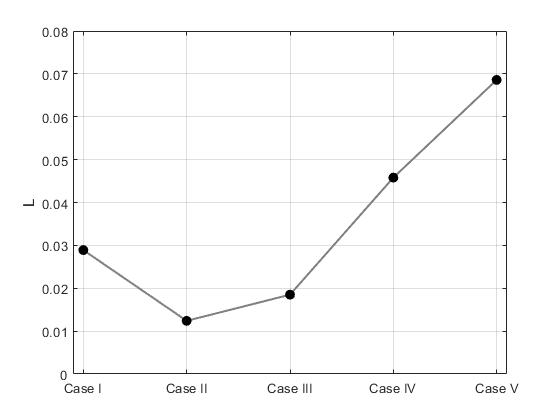}
    \caption{Change in expected loss when input nodes change, for different configurations of entry nodes.}
    \vspace{-0.1in}
    \label{fig: plot_inp_nodes}
\end{figure}

We begin by looking at how the entry nodes affect the optimal investment profile and the expected loss. {We consider the following scenarios: (Case I) the base configuration in Figure~\ref{reduced_attack_graph}; (Case II) no mobile communication by removing the CELL and TELE nodes; (Case III) no physical access by removing the USB node; (Case IV)  one additional entry node with low standalone loss; and (Case V) one additional entry node with high standalone loss.} 
Figures \ref{fig: bar_inp_nodes} and \ref{fig: plot_inp_nodes} depict the variation in the investment on nodes and optimal loss for each case. 

Cases II and III have a smaller attack surface (number of entry nodes) compared to Cases IV and V. This translates to a higher expected loss compared to $\mathbf{L_{\text{base}}}$, as the limited security resources must now be stretched across a higher number of inputs, with each input receiving a lower investment. This matches with the expectation that a ``smart'' vehicle with many connectivity elements is more vulnerable.

We further observe that there two distinct features of the added nodes that impact how much they increase the expected losses by demanding a higher share of the limited security budget: (1) an input node with higher accessibility (low $p^0$ in our model) demands a higher investment (compare Case II: removing the CELL node with Case III: removing the USB node), and (2) a more sophisticated added node which has higher standalone losses (higher $L$ in our model), demands higher investments (compare Case IV with Case V). 
A node with any of these two characteristics will be more detrimental to the overall vehicle security. 
This indicates that while entry nodes may be beneficial for connectivity reasons, having them perform even minor safety-critical functions may not be desirable. 

While protecting/``closing'' these entry points may be one of the solutions to improve the vehicle's security, in many cases the presence of these nodes is dictated by factors beyond the control of the security designers. In such cases it becomes important to introduce security measures downstream of the graph to improve its security. Such measures show up as series nodes on the attack graph as they are an additional item to compromise for the attacker. We consider these design interventions next. 

\subsubsection{Addition of downstream defensive measures} 
\begin{figure}[t]
    \centering
    \captionsetup{justification=centering}
    \includegraphics[width=0.64\columnwidth]{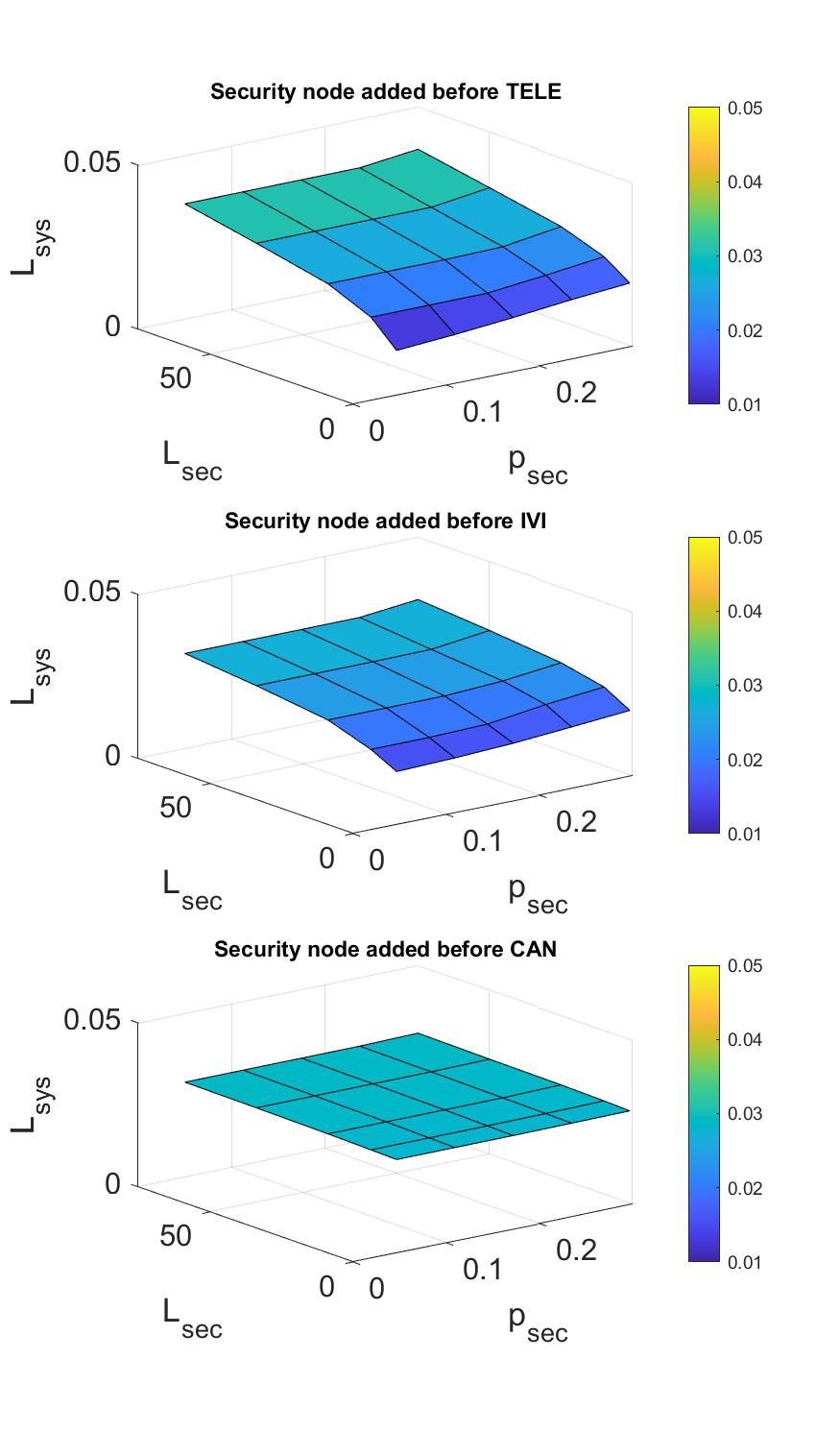}
    \vspace{-0.1in}
    \caption{Change in expected loss when a defensive measure is added in different locations in the network.}
    \vspace{-0.1in}
    \label{fig: plot_sec_nodes}
\end{figure}

We consider defense measures such as the introduction of trust zones, encrypting messages, secure memory access, and the like.  The probability of attack on these security mechanisms, $p_{\text{sec}}$ and their corresponding standalone loss, ${L}_{\text{sec}}$, depends on the mechanisms introduced. Figure \ref{fig: plot_sec_nodes} shows the effect of introducing such nodes at three different locations on $\mathcal{G}$: before the TELE node, before the IVI node, and before the CAN node. Further, we consider the  effect of varying the security characteristics of the introduced measures. 

As expected, security mechanisms which require physical access or high user interaction have a lower $p_{\text{sec}}$ and generally reduce $\mathbf{L}_{\text{sys}}$. However, this effect becomes negligible when the safeguards are introduced closer to the final target: it can be seen that the expected loss for this system $\mathbf{L}_\text{sys}$ is mostly flat when the security measure is introduced before the CAN. In other words, security measures implemented further upstream can influence the loss much more than those introduced closer to the target; this is consistent with the tendency to encourage perimeter defenses (e.g. firewalls). 
Further, we note that endowing the added security mechanism with additional safety-critical responsibilities (higher ${L}_{\text{sec}}$) may not be advisable. This can be seen from Fig. \ref{fig: plot_sec_nodes} where a high ${L}_{\text{sec}}$ leads to $\mathbf{L}_{\text{sys}} \geq \mathbf{L}_{\text{base}}$ irrespective of $p_{\text{sec}}$. This argues in favour of implementation of passive security measures vs. active measures which take on additional functionalities or interact with other assets in the system.

\subsubsection{Functional (information) redundancy for better security} The two types of interventions we have considered so far (modifying entry nodes, and adding downstream safeguards) have called for direct redesign of the Infotainment Module. However, typically, these modules are acquired from third-party suppliers, and guaranteeing security across the supply chain is hard for the OEM. 
In such a case, it is up to the OEM to design the vehicle's internal architecture to be resilient to security attacks. We next consider such choices. 

\begin{figure}[t]
    \centering
    \captionsetup{justification=centering}
    \includegraphics[width=0.8\columnwidth]{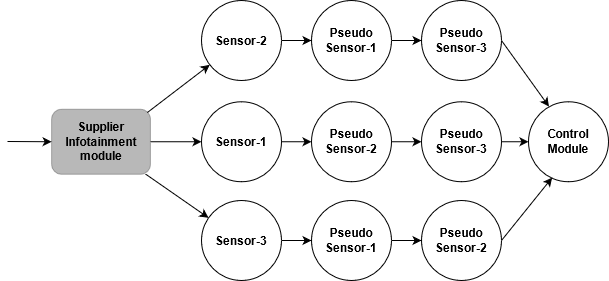}
    \caption{Attack graph with improved security through introduction of functional redundancy}
    \label{fig: par_sec_graph}
\end{figure}

Consider a remote attack launched \textit{via.} the Infotainment Module with the goal of sending malicious sensor messages over the vehicles internal network. 
Such message integrity attacks can cause safety hazards if the messages are used for decision-making or control synthesis in safety-critical functionalities.
In this case, the designer can use functionally redundant sensors to improve the security status of the system (shown in Figure \ref{fig: par_sec_graph}). 
As an example, this modified attack graph could reflect that some state information, such as the vehicle's speed, is obtained from various sources such as GPS, wheel speed sensors, and engine speed, with all the information used by the control module. 
Such functional redundancy forces the attacker to compromise all sensors to gain access to the control module undetected. 

Table \ref{table: redundant_sensors} compares the expected loss as the redundancy in the system is increased. Firstly, we note the choice of different probabilities of attack for different sensors. This is intended to capture differences in the attack requirements. For example, a sensor may be hard-wired to the control module and require physical access for measurement manipulation. Such sensor's measurements are hardest to manipulate considering its physical attack vector. We notice that in the case when a single redundant sensor is introduced, although its attack probability is higher, it improves the overall security of the system. With further increase in redundancy, the expected loss decreases. Note that the rate of this decrease is itself decreasing (i.e. the first redundant sensor improves security more than any subsequent sensors). 

\begin{table}[t]
\caption{Impacts of adding redundant sensors}
\label{table: redundant_sensors}
\vspace{-0.1in}
\begin{center}
\begin{tabular}{|p{1in}||p{1.4in}||c|}
\hline
Description & Sensor attack  probability & $\mathbf{L_{arch}}$ \\
& & \\
\hline
Base configuration & 0.207 & 0.0125  \\
\hline
1 redundant sensor & (0.207,0.252) & 0.0121 \\
\hline
2 redundant sensors & (0.207,0.252,0.144) & 0.0120 \\
\hline
3 redundant sensors & (0.207,0.252, 0.144, 0.144) & 0.0119 \\
\hline
\end{tabular}
\vspace{-0.2in}
\end{center}
\end{table}

\section{Conclusion and Future Work}\label{sec:conclusion}

We studied how network design interventions impact the security of a CPS network modeled by an attack graph. 
Specifically, we considered four alternatives for introducing new nodes in the system: in series  (added safeguards), in parallel  (structural redundancy), a hybrid of series and parallel connections (functional redundancy), and as an additional input (added functionalities). We identified their impacts on the overall network security, by analyzing the changes in the defender's optimal investment strategies following the introduction of these nodes, and the changes in the adversary's attack strategies and inflicted losses. 
Finally, we showed how this approach could help guide network re-design strategies on a simple automotive application with practical design restrictions. Future research avenues include considering a limited budget for the attacker and the study of other attacker behaviour models.

\bibliographystyle{IEEEtran}
\bibliography{ref.bib}

\newpage
\appendix
\section{APPENDIX}

\section*{Proof of Lemma~\ref{lemma:series_loss_modification}}

\begin{proof}
We begin by showing that in the optimal investment profile $\mathbf{x}^*$, $x^*_i\geq 0$ and $x^*_{j}=0, \forall j=i+1, \ldots, i+k$. 

First, note that the loss over this stretch of assets under a profile of investments $\mathbf{x}$ is given by:
\begin{align*}
    \mathbf{L}_{i\rightarrow i+k}(\mathbf{x}) := \sum_{l=i}^{i+k} L_{l} (\prod_{t=i}^l p^0_t e^{-x_t}) = \sum_{l=i}^{i+k} L_{l} e^{-\sum_{t=i}^l x_t}\prod_{t=i}^l p^0_t
\end{align*}
Assume in contradiction that in the optimal investment profile $x_j^*>0$ for some $j\neq i$. Consider an alternative profile $\mathbf{\tilde{x}}$ built from $\mathbf{x}^*$ where this investment is shifted to $v_i$, i.e. that $\tilde{x}^i=x_i^*+x_j^*$ and $\tilde{x}_j=0$. 
Then, 
\begin{align*}
    & \mathbf{L}_{i\rightarrow i+k}(\mathbf{\tilde{x}}) - \mathbf{L}_{i\rightarrow i+k}(\mathbf{x}^*) \\
    &= \sum_{l=i}^{i+k} L_{l} e^{-\sum_{t=i}^l \tilde{x}_t}\prod_{t=i}^l p^0_t - \sum_{l=i}^{i+k} L_{l} e^{-\sum_{t=i}^l x^*_t}\prod_{t=i}^l p^0_t\\
    &= \sum_{l=i}^{i+k} L_{l} \prod_{t=i}^l p^0_t \left(e^{-\sum_{t=i}^l \tilde{x}_t} - e^{-\sum_{t=i}^l x^*_t} \right) < 0~.
\end{align*}
The last line follow from the fact that $\sum_{t=i}^l \tilde{x}_t > \sum_{t=i}^l {x}^*_t$ for $l<j$, and $\sum_{t=i}^l \tilde{x}_t = \sum_{t=i}^l {x}^*_t$ for $l\geq j$. Therefore, $\mathbf{\tilde{x}}$ achieves a lower cost than the optimal $\mathbf{x}^*$, which is a contradiction.

We conclude that any investments in the optimal profile for $\mathcal{G}$ will only be placed on node $v_i$. Therefore, ignoring all remaining nodes $\{v_{i+1} \rightarrow v_{i+2} \rightarrow \ldots \rightarrow v_{i+k}\}$, graphs $\mathcal{G}$ and $\mathcal{G}'$ are the same, and therefore have the same optimal investments profile $\mathbf{x}^*$. 

Further, in $\mathcal{G}$, the effect of the stretch of nodes $\{v_i \rightarrow v_{i+1} \rightarrow \ldots \rightarrow v_{i+k}\}$ on the optimal cost is given by
\begin{align*}
    \mathbf{L}_{i\rightarrow i+k}(\mathbf{x}^*) = \sum_{l=i}^{i+k} L_{l} e^{-\sum_{t=i}^l x_t}\prod_{t=i}^l p^0_t = e^{-x_i^*} \sum_{l=i}^{i+k} L_{l} \prod_{t=i}^l p^0_t.
\end{align*}
This is the same as the loss in $\mathcal{G}'$ under profile $\mathbf{x}^*$. Therefore, the two attack graphs have the same loss under and the same investment profile, and are therefore equivalent. 
\end{proof}

\section*{Proof of Lemma~\ref{lemma:parallel_loss_modification}}

\begin{proof}
We begin by showing that in the optimal investment profile $\mathbf{x}^*$, $x_i^* \geq 0$, and $x_j^*=0$, $\forall v_j \in \mathcal{P}_{p, i\rightarrow k}, j\neq i$. 
Assume, by contradiction, that there is a node $v_j$ in one of the parallel paths, $P_j\in \mathcal{P}_{p, i\rightarrow k}$ such that $x_j^*>0$. Define an alternative profile $\mathbf{\tilde{x}}$ that is the same as $\mathbf{x}^*$, except that the investment on $v_j$ is shifted to $v_i$, i.e. that $\tilde{x}^i=x_i^*+x_j^*$ and $\tilde{x}_j=0$. Consider the two following cases:
\begin{enumerate}
    \item $P_j$ is the path with the highest loss among all parallel paths $\mathcal{P}_{p, i\rightarrow k}$ under the investment profile $\mathbf{x}^*$: Then, by the same arguments as in Lemma~\ref{lemma:series_loss_modification}, the alternative profile $\mathbf{\tilde{x}}$ would achieve a lower loss than $\mathbf{x}^*$, which would be a contradiction.  
    \item There is a $P_l\in \mathcal{P}_{p, i\rightarrow k}$ with a higher loss than $P_j$ under the investment profile $\mathbf{x}^*$: Then, shifting the investment from $v_j$ to the common origin node $v_i$, as done in $\mathbf{\tilde{x}}$, would increase the investments on $P_l$ and reduce its loss,  contradicting the optimality of $\mathbf{x}^*$. 
\end{enumerate}
This means that in an optimal investment profile, security investments (if any) will only be placed on the common origin node of the paths in $\mathcal{P}_{p, i\rightarrow k}$.

We conclude that all paths in $\mathcal{P}_{p, i\rightarrow k}$ have the same total investment on them at the optimal solution. Therefore, if the attacker's selected worst-case attack path passes through one of the $P\in \mathcal{P}_{p, i\rightarrow k}$, it will be the one which has the higher compound loss $\sum_{v_l\in P} L_{l} \prod_{v_t\in Pre(v_l)} p^0_t$ and would therefore maximize the attacker's payoff. All other paths can be ignored in the evaluation of the optimal cost of $\mathcal{G}$, and this will be the same as the optimal cost on $\mathcal{G}'$ under $\mathbf{x}^*$. 
\end{proof}

\section*{Additional details on the derivation of the game parameters}

Tables~\ref{ae} and \ref{im} provide additional details on the derivation of the game parameters in Table~\ref{table: p0_nodes} for our numerical experiments. 

\begin{table}[thbp!]
\caption{Attack ease metrics based on CVSSv3}
\label{ae}
\begin{center}
\begin{tabular}{|c||c||c|}
\hline
Metric & Rank & Value \\
\hline
Attack Vector (AV) & Network (N) & 0.85 \\
& Adjacent (A) & 0.62 \\
& Local (L) & 0.55 \\
& Physical (P) & 0.2 \\
\hline
Attack Complexity (AC) & Low (L) & 0.77 \\
& High (H) & 0.44 \\
\hline
Privileges Required (PR) & None (N) & 0.85 \\
& Low (L) & 0.62 (scope unchanged) \\
& High (H) & 0.27 (scope unchanged) \\
\hline
User Interaction (UI) & High (H) & 0.85 \\
& Required (R) & 0.62 \\
\hline
\multicolumn{3}{|c|}{Exploitability (EX) = 8.22 x AV x AC x PR x UI} \\
\hline
\end{tabular}
\end{center}
\end{table}

\begin{table}[thbp!]
\caption{Impact metrics based on CVSSv3}
\label{im}
\begin{center}
\begin{tabular}{|c||c||c|}
\hline
Metric & Rank & Value \\
\hline
Impact-Confidentiality (ImC)/ & High (H) & 0.56 \\
Impact-Integrity (ImI)/ & Low (L) & 0.22 \\
Impact-Availability (ImA) & None (N) & 0 \\
\hline
\multicolumn{3}{|c|}{Impact (Im) = 6.42 x [1 - ( (1-ImC) x (1-ImI) x (1-ImA) )]} \\
\hline
\end{tabular}
\end{center}
\end{table}

\end{document}